\documentclass[aps,prb,twocolumn,preprintnumbers,amsmath,amssymb,superscriptaddress]{revtex4}%

\usepackage{graphicx}%
\usepackage{dcolumn}
\usepackage{amsmath}
\usepackage{color}

\newcommand{\RN}[1]{\textup{\uppercase\expandafter{\romannumeral#1}}}%

\begin{document}

\title{Structural phases of elemental Gallium: universal relations in type-I superconductors}

\author{Rustem~Khasanov}
 \email{rustem.khasanov@psi.ch}
 \affiliation{Laboratory for Muon Spin Spectroscopy, Paul Scherrer Institute, CH-5232 Villigen PSI, Switzerland}

\author{Hubertus Luetkens}
 \affiliation{Laboratory for Muon Spin Spectroscopy, Paul Scherrer Institute, CH-5232 Villigen PSI, Switzerland}

\author{Alex Amato}
 \affiliation{Laboratory for Muon Spin Spectroscopy, Paul Scherrer Institute, CH-5232 Villigen PSI, Switzerland}
 
\author{Elvezio Morenzoni}
\email{elvezio.morenzoni@psi.ch}
 \affiliation{Laboratory for Muon Spin Spectroscopy, Paul Scherrer Institute, CH-5232 Villigen PSI, Switzerland}

\begin{abstract}

The temperature dependent measurements of the thermodynamic critical field  and the specific heat  for the pressure stabilized Ga-II phase of elemental Gallium are presented. The discussion of these and other Ga phases data in the context of elemental and binary phonon-mediated \mbox{type-I} superconductors allowed to establish simple scaling relations between BCS quantities such as $B_{\rm c}(0)/T_{\rm c}\sqrt{\gamma_{\rm e}}$ and the specific heat jump at $T_{\rm c}$
versus the coupling strength $2\Delta/k_{\rm B} T_{\rm c}$ [$\Delta$ and $B_{\rm c}(0)$ are the zero-temperature values of the superconducting energy gap and the thermodynamic critical field, respectively, $T_{\rm c}$ is the transition temperature, and $\gamma_{\rm e}$ is the electronic specific heat]. The scaling relations can be analytically expressed by taking into account strong-coupling corrections to BCS theory.
Such correlations can naturally explain the linear relation between $B_{\rm c}(0)$ and $T_{\rm c}$, which holds for type-I superconducting materials.

\end{abstract}


\maketitle


Understanding the phenomenon of superconductivity, which is observed in quite different systems, such as metallic elements, molecular metals, Weyl and Dirac semimetals, cuprates, pnictides,  {\it etc.} involves searching for commonalities and general relations between fundamental physical parameters across various materials. The existence of such universal trends  may point to common underlying pairing mechanisms.
For unconventional type-II superconductors one of the earliest reported trend was the linear scaling between the superconducting transition temperature ($T_{\rm c}$) and the superfluid density ($\rho_{\rm s}$), first identified by Uemura {\it et al.}\cite{Uemura_PRL_1989,Uemura_PRL_1991} for hole doped cuprate superconductors. Similar linear relations were further observed  in electron doped cuprates,\cite{Shengelaya_PRL_2005, Khasanov_PRB_2008_InfLayer} Fe-based superconductors,
\cite{Luetkens_PRL_2008,Khasanov_PRB_2008,Goko_PRB_2009,Pratt_PRB_2009,Khasanov_PRL_2010} layered Weyl,\cite{Guguchia_NatComm_2017} and Dirac semimetals.\cite{Guguchia_Arxiv_2019} In molecular superconductors  $\rho_{\rm s}$ was found to
be proportional to $T_{\rm c}^{2/3}$, \cite{Pratt_PRL_2005} while in some phonon mediated BCS superconductors $\rho_{\rm s}\propto
T_{\rm c}^{3}$.\cite{Khasanov_PRB-NbB2_2008, Khasanov_SciRep_2015}
Later on a scaling relations between $\rho_s$ and the dc conductivity $\sigma_{\rm dc}$ was
suggested and a linear relation between $\rho_{\rm s}$ and the product $\sigma_{\rm dc} T_{\rm c}$ was demonstrated in Refs.~\onlinecite{Homes_Nature_2004, Zaanen_Nature_2004, Dordevic_SciRep_2013} for a set of cuprates, pnictides and molecular superconductors. In addition to these, there are many other types of relations for various classes of type-II superconductors mentioned in the literature (see {\it e.g.} Refs.~\onlinecite{Budko_PRB_2009, Kogan_PRB_2009, Zaanen_PRB_2009, Kim_JPCM_2011, Stewart_RMP_2011, Kim_PRB_2012, Budko_PRB_2014, Kim_PRB_2015, Liu_arxiv_2019} and references therein).

\begin{figure}[htb]
\includegraphics[width=0.78\linewidth]{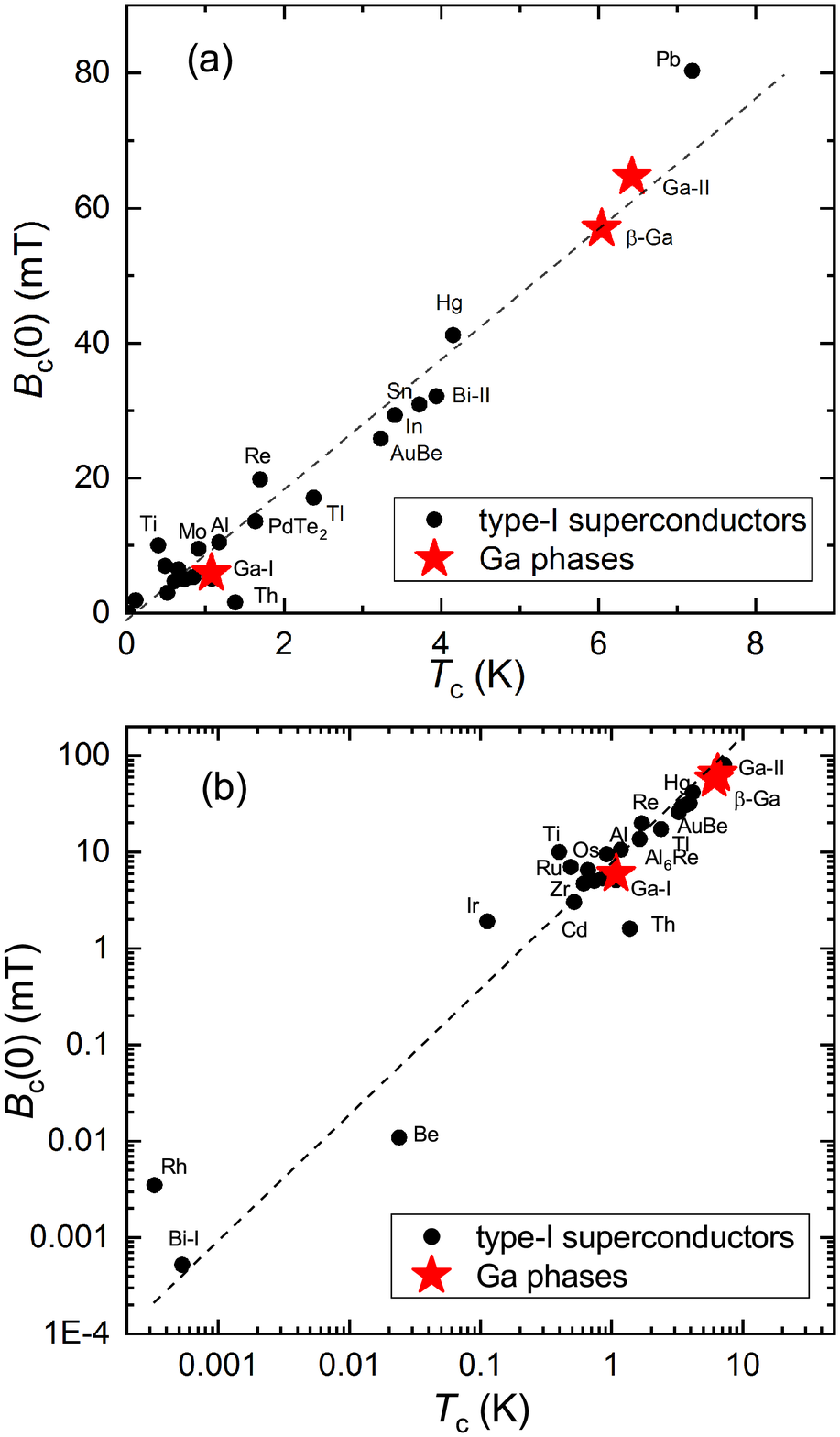}
\caption{ (a) Empirical relation between the zero-temperature value of the thermodynamic critical field $B_{\rm c}(0)$ and the transition temperature $T_{\rm c}$ for type-I superconductors, after Rohlf, Ref.~\onlinecite{Rohlf_Book_1994}. (b) Same as in panel (a) but in $log-log$ scale. Black symbols are data points from Refs.~\onlinecite{Kittel_Book_1996, Poole_Book_2014, Leng_PRB_2017, Prakash_Science_2017, Campanini_PRB_2018, Singh_PRB_2019, Beare_PRB_2019, Khasanov_Arxiv_2019, Peets_PRB_2019}. Red stars correspond to different Gallium phases (see Table~\ref{tab1}).}
 \label{fig:Bc-vs-Tc}
\end{figure}

For conventional phonon-mediated type-I superconductors, the microscopic theories such as BCS,\cite{Bardeen} or the more general Eliashberg  theory,\cite{Eliashberg_1960a,Eliashberg_1960b, Carbotte_RMP_1990, Marsiglio_book_2008, Scalapino_book_1969, McMillan_book_1969, Allen_book_1982} accounting for retardation effects, were very successful in describing the observed properties. Empirical scaling relation between the zero-temperature value of the thermodynamic critical field $B_{\rm c}(0)$ and the transition temperature $T_{\rm c}$ was established,\cite{Rohlf_Book_1994} but it receives much less attention than the mentioned above universal trends of type-II superconducting materials.\cite{Uemura_PRL_1989,Uemura_PRL_1991, Shengelaya_PRL_2005, Khasanov_PRB_2008_InfLayer, Luetkens_PRL_2008,Khasanov_PRB_2008,Goko_PRB_2009,Pratt_PRB_2009,Khasanov_PRL_2010, Guguchia_NatComm_2017, Guguchia_Arxiv_2019, Pratt_PRL_2005, Khasanov_PRB-NbB2_2008, Khasanov_SciRep_2015, Homes_Nature_2004, Zaanen_Nature_2004, Dordevic_SciRep_2013, Budko_PRB_2009, Kogan_PRB_2009, Zaanen_PRB_2009, Kim_JPCM_2011, Stewart_RMP_2011, Kim_PRB_2012, Budko_PRB_2014, Kim_PRB_2015, Liu_arxiv_2019}  The panels (a) and (b) of Fig.~\ref{fig:Bc-vs-Tc} display such a  $B_{\rm c}(0)$ {\it vs.} $T_{\rm c}$ relation using a linear and a $log-log$ scale, respectively (the data points are taken from Refs.~\onlinecite{Kittel_Book_1996, Poole_Book_2014, Leng_PRB_2017, Prakash_Science_2017, Campanini_PRB_2018, Singh_PRB_2019, Beare_PRB_2019, Khasanov_Arxiv_2019, Peets_PRB_2019}) and show that a simple linear proportionality holds well over a wide temperature range.
The fact, that the critical magnetic field required to destroy the superconducting state is strongly correlated with the critical temperature, indicates that both quantities [$T_{\rm c}$ and $B_{\rm c}(0)$] are a measure of the energy, which has to be supplied to the material to destroy superconductivity. This is consistent with the presence of a bandgap between the superconducting and normal state.

The direct relationship between $T_{\rm c}$ and the zero temperature value of the superconducting energy gap $\Delta$ is a result of BCS theory predicting a dimensionless ratio $\Delta/k_{\rm B} T_{\rm c}=1.764$.\cite{Bardeen}
For superconductors not in the weak coupling limit this ratio is not universal and its effective value $\alpha=\Delta/k_{\rm B} T_{\rm c}$  is generally taken as a measure of the coupling character of the material. In single gap phonon mediated superconductors $\alpha$ typically lies in the range of $1.76\lesssim \alpha \lesssim 2.5$.\cite{Carbotte_RMP_1990}  The parameter $\alpha$ becomes a central quantity of the so-called $\alpha-$model, a widely used adaption of BCS theory, where the proportionality factor between $T_{\rm c}$ and $\Delta$ is taken as an adjustable parameter to describe electronic and thermodynamic properties of the superconductor. \cite{Padamsee_JLTP_1973,Johnston_SST_2013}

Note, however, that while the proportionality between $T_{\rm c}$ and $\Delta$ is a well accepted fact, it is not evident why $B_{\rm c}(0)$ should follow a similar trend and what governs the proportionality factor between $B_{\rm c}(0)$ and $\Delta$, as well as between $B_{\rm c}(0)$ and $T_{\rm c}$.
Additionally, from the experimental point of view, there is a lack of points for type-I superconductors with high transition temperature  values. The gap between Hg ($T_{\rm c}\simeq 4.2$~K) and the next representative of 'high-$T_{\rm c}$' type-I superconductor Pb ($T_{\rm c}\simeq 7.2$~K) is almost a factor of 2 [see the black points Fig.~\ref{fig:Bc-vs-Tc}~(a)].

This paper addresses both above mentioned points. It was shown, in particular, that in phonon-mediated superconductors there are simple correlations between the coupling strength $2\Delta/k_{\rm B} T_{\rm c}$, the ratio $B_{\rm c}(0)/T_{\rm c}\sqrt{\gamma_{\rm e}}$ ($\gamma_{\rm e}$ is the electronic specific heat), and the specific heat jump at $T_{\rm c}$, $\Delta C(T_{\rm c})/\gamma_{\rm e} T_{\rm c}$. Such correlations can be well explained by considering strong coupling corrections to the relevant quantities and can naturally explain the linear relation between $B_{\rm c}(0)$ and $T_{\rm c}$ as well give an estimate of the $B_{\rm c}(0)/T_{\rm c}$ ratio.
To address the lack of experimental points between Hg and Pb, the temperature dependence of the thermodynamic critical field $B_{\rm c}(T)$
of the high-pressure structural phase of elemental Gallium, the so called Ga-II phase ($T_{\rm c}\simeq 6.5$~K, Refs.~\onlinecite{Buckel_ZfP_1963, Eichler_ZPhysB_1980}), was studied. In addition, the available specific heat data of Ga-II phase reported in Ref.~\onlinecite{Eichler_ZPhysB_1980} were reevaluated.

The temperature dependence of the thermodynamic critical field $B_{\rm c}(T)$ of Ga-II was determined in transverse-field muon-spin rotation (TF-$\mu$SR) measurements.   Experiments were performed at the $\mu$E1 beam line (Paul Scherrer Institute, Switzerland) by using the GPD (General Purpose Decay) $\mu$SR spectrometer.\cite{Khasanov_HPR_2016} As a local magnetic probe, $\mu$SR is well suited to study type-I superconducting materials (see {\it e.g.} Ref.~\onlinecite{Karl_Arxiv_2019} and references therein). 
The measurements were performed on a commercial Ga specimen of $99.999\%$ purity. The sample was casted into a $\simeq5.9$~mm in diameter and $\simeq10.5$~mm in height cylinder and placed inside a double-wall piston-cylinder pressure cell made of MP35N/NiCrAl alloy. The pressure cell is similar to the one described in Refs.~\onlinecite{Khasanov_HPR_2016,Shermadini_HPR_2017}. The pressure necessary to establish the Ga-II phase ($p\simeq 2.48$~GPa) was determined by measuring the superconducting transition of a small piece of In (pressure indicator) placed inside the cell together with the Ga sample. Details of TF-$\mu$SR under pressure measurements of Ga-II will be published separately.

The temperature dependence of the difference between the superconducting state ($C_{\rm S}$) and the normal state ($C_{\rm N}$) specific heat components [$(C_{\rm S}-C_{\rm N})/T$ {\it vs.} $T$] of a Ga-II superconductor were evaluated from the measurements of Eichler {\it et al.}\cite{Eichler_ZPhysB_1980} Experiments in Ref.~\onlinecite{Eichler_ZPhysB_1980} were performed  at $p\simeq 3.5$~GPa. The $C_{\rm S}/T$ and $C_{\rm N}/T$ data were corrected by taking into account the pressure induced molar volume change, which is associated with the Ga-I to Ga-II structural phase transition.\cite{Degtyareva_PRL_2004, Lyapin_ZhETP_2008} Note that without such correction the value of the electronic specific heat $\gamma_{\rm e}$, as is reported in Ref.~\onlinecite{Eichler_ZPhysB_1980}, is underestimated.

\begin{figure}[htb]
\includegraphics[width=0.8\linewidth]{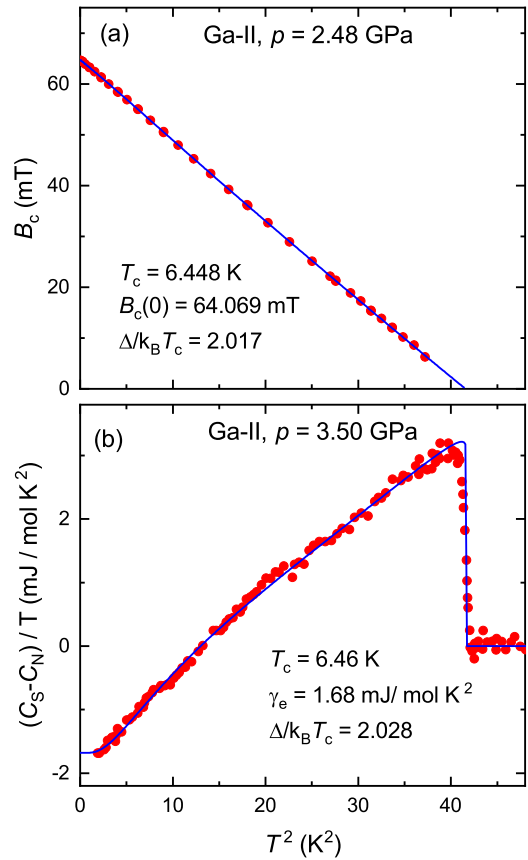}
\caption{ (a) $T^2$ dependence of the thermodynamic critical field $B_{\rm c}$ of Ga-II at $p\simeq 2.48$~GPa as obtained in our TF-$\mu$SR experiment. (b) $T^2$ dependence of the difference between the superconducting state ($C_{\rm S}/T$) and the normal state ($C_{\rm N}/T$) specific heat components of Ga-II, after Ref.~\onlinecite{Eichler_ZPhysB_1980}. The $C_{\rm S}/T$ and $C_{\rm N}/T$ data from Ref.~\onlinecite{Eichler_ZPhysB_1980} were corrected for  the pressure induced molar volume change, which is associated with the Ga-I to Ga-II structural phase transition.\cite{Degtyareva_PRL_2004,Lyapin_ZhETP_2008} Blue lines are fits of the $\alpha-$model (see text for details).}
 \label{fig:Bc-and-Cp}
\end{figure}

The experimental $B_{\rm c}$ {\it vs.} $T^2$ and $[C_{\rm S}-C_{\rm N}]/T$ {\it vs.} $T^2$ curves are shown in panels (a) and (b) of Fig.~\ref{fig:Bc-and-Cp}.
The data were analyzed within the framework of the phenomenological $\alpha-$model.\cite{Padamsee_JLTP_1973, Johnston_SST_2013} This model assumes, on one hand,  that the temperature dependence of the normalized superconducting energy gap, ${\Delta(T)}/{\Delta}={\Delta_{\rm BCS}(T)}/{\Delta_{\rm BCS}}$, is that given by BCS theory. \cite{Muehlschlegel_ZPhys_1959}
On the other hand, to calculate the temperature dependence of the electronic free energy, entropy, heat capacity and thermodynamic critical field, the $\alpha-$model allows $\alpha=\Delta/k_{\rm B} T_{\rm c}$ to be an adjustable parameter.
The thermodynamic critical field $B_{\rm c}(T)$ and the specific heat jump $\Delta C(T_{\rm c})/\gamma_{\rm e} T_{\rm c}$ can be calculated within this model. The relevant expression are given in Refs.~\onlinecite{Padamsee_JLTP_1973,Johnston_SST_2013}.
Fits of the $\alpha-$model to the $B_{\rm c}(T)$ and $\Delta C/T (T)$ data are shown by blue lines in Fig.~\ref{fig:Bc-and-Cp}. The fit results for Ga-II, as well as the corresponding values for the other superconducting phases of  elemental Gallium are summarized in Table~\ref{tab1}.

\begin{table*}[htb]
\caption{\label{tab1} Parameters for different Gallium phases. $T_{\rm c}$ is the superconducting transition temperature,  $B_{\rm c}(0)$ is the zero-temperature value of thermodynamical critical field, $\alpha=\Delta/k_{\rm B} T_{\rm c}$ is the coupling strengths, $\gamma_{\rm e}$ is the electronic specific heat coefficient,  and $\Delta C (T_{\rm c}) /\gamma_{\rm e}T_{\rm c}$ is the specific heat jump at $T_{\rm c}$.   }
\begin{tabular}{ccccccccccc}
\hline
\hline
Ga phases&Pressure&Superconductivity&$T_{\rm c}$&$B_{\rm c}(0)$&$\Delta/k_{\rm B} T_{\rm c}$&$\gamma_{\rm e}$& $\Delta C (T_{\rm c}) /\gamma_{\rm e}T_{\rm c}$& References\\
&&&(K)&(mT)&& (${\rm mJ}/{\rm mol}\;{\rm K}^{2}$)&&\\
\hline
Ga-I & 1~bar    &type-I&1.08 &5.03  &$\sim$1.75&0.600 &1.41 &Refs.~\onlinecite{Seidel_PR_1958, Gregory_PR_1966, Phillips_PR_1964, Goodman_PhilMag_1951, Yoshihiro_JPSJ_1970}\\
$\beta-$Ga&1~bar&type-I&6.04 &57    &2.00      &1.53  &1.83 &Ref.~\onlinecite{Campanini_PRB_2018}\\
Ga-II &2.48~GPa &type-I&6.448&64.069&2.017     &--    & --  &This work\\
Ga-II & 3.5 GPa &type-I&6.46 &  --  &2.028     &1.68  &2.028&This work and Ref.~\onlinecite{Eichler_ZPhysB_1980}\\
\hline
\hline
\end{tabular}
\end{table*}

\begin{figure}[htb]
\includegraphics[width=0.8\linewidth]{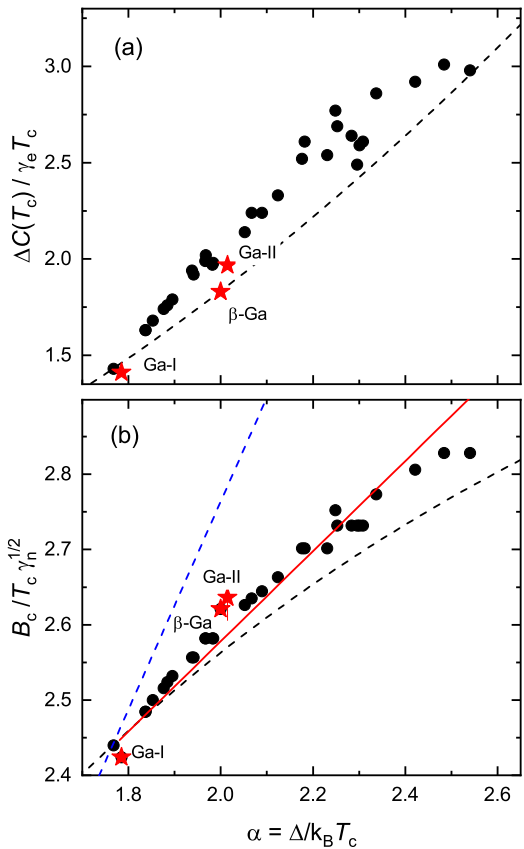}
\caption{ (a) Specific heat jump at the superconducting transition temperature $\Delta C (T_{\rm c}) /\gamma_{\rm e}T_{\rm c}$ and
(b) ratio $B_{\rm c}(0)/T_{\rm c}\sqrt{\gamma_{\rm e}}$ versus the coupling strength $2\Delta/k_{\rm B}T_{\rm c}$
for various elemental and binary phonon-mediated superconductors.
Red stars correspond to the structural phases of elemental Gallium (see Table~\ref{tab1}). Black dots are values from Tables~IV and X of Ref.~\onlinecite{Carbotte_RMP_1990}.
Dashed lines are predictions of the phenomenological $\alpha-$model (see text for details). The red line is the theoretical prediction of Eq.~\ref{eq:Bc_relation}.}
 \label{fig:Relations_alpha-model}
\end{figure}

As a further step, the physical parameters of the various phases of Gallium, as summarised in Table ~\ref{tab1}, were compared with those reported in the literature for various phonon-mediated type-I superconductors. The values for elemental and binary superconductors were taken from Tables~IV and X of Ref.~\onlinecite{Carbotte_RMP_1990}.
Figure~\ref{fig:Relations_alpha-model} plots the specific heat jump at the superconducting transition temperature $\Delta C (T_{\rm c}) /\gamma_{\rm e}T_{\rm c}$ [panel (a)] and the dimensionsless parameters $B_{\rm c}(0)/T_{\rm c}\sqrt{\gamma_{\rm e}}$ [panel (b)] versus the coupling strength $\alpha=\Delta/k_{\rm B} T_{\rm c}$.
Note that the selection of plot parameters is a natural consequence of the BCS theory and of the $\alpha-$model. BCS theory predicts well defined values for all these quantities:
$\alpha_{\rm BCS}=1.764$, $B_{\rm c}(0)/T_{\rm c}\sqrt{\gamma_{\rm e}} = 2.438$ and $\Delta C (T_{\rm c}) /\gamma_{\rm e}T_{\rm c} = 1.426$,\cite{Bardeen} while  the $\alpha-$model allows to express the latter quantities as a function of $\alpha$.

From the data presented in Fig.~\ref{fig:Relations_alpha-model} two important points emerge: \\
(i) The experimental data for the various Gallium phases (denoted by red stars) are in a good agreement with those of the other type-I superconductors.   \\
(ii) Both quantities,  $\Delta C (T_{\rm c}) /\gamma_{\rm e}T_{\rm c}$ and $B_{\rm c}(0)/T_{\rm c}\sqrt{\gamma_{\rm e}}$, directly scale with the coupling strengths $\Delta/k_{\rm B}T_{\rm c}$.
To the best of our knowledge, these experimental scalings between BCS parameters of phonon-mediated superconductors have not been discussed before.

The predictions of the $\alpha-$model are represented by dashed lines in panels (a) and (b) of Fig.~\ref{fig:Relations_alpha-model}.
In the following the quantities $\Delta C (T_{\rm c}) /\gamma_{\rm e}T_{\rm c}$ and $B_{\rm c}(0)/T_{\rm c}\sqrt{\gamma_{\rm e}}$ are going to be discussed separately. For the specific heat jump an analytical expression given in Ref.~\onlinecite{Johnston_SST_2013} reads:
\begin{equation}
\frac{\Delta C(T_{\rm c})}{\gamma_{\rm e} T_{\rm c}}\approx 1.426 \left(\frac{\alpha}{\alpha_{\rm BCS}} \right)^2 \approx 0.458 \; \alpha^2.
 \label{eq:Cp_relation_alpha-model}
\end{equation}
The quadratic dependence of $\Delta C (T_{\rm c}) /\gamma_{\rm e}T_{\rm c}$ on $\alpha$ was also remarked in Ref.~\onlinecite{Bouquet_EPL_2001}.
Figure~\ref{fig:Relations_alpha-model}~(b) shows that the experimental values lie slightly higher than it is expected from the $\alpha-$model. However, the overall agreement (within only 5-10\%) is quite good. A power law fit yields parameters quite close to the theoretical prediction: ${\Delta C(T_{\rm c})}/{\gamma_{\rm e} T_{\rm c}}\approx 0.469 \; \alpha^{2.10}$.

Figure~\ref{fig:Relations_alpha-model}~(b) compares the experimental dependence of $B_{\rm c}(0)/T_{\rm c}\sqrt{\gamma_{\rm e}}$ on $\Delta/k_{\rm B} T_{\rm c}$ with $\alpha-$model predictions.
In principle, $B_{\rm c}(T)$ can be determined from the free energy difference between normal and superconducting state (at zero applied field) or from the difference of the corresponding entropies.\cite{Padamsee_JLTP_1973, Johnston_SST_2013}
Surprisingly, the two approached give the same result only for  $\alpha\equiv\alpha_{\rm BCS}=1.764$ where
$B_{\rm c}(0)/T_{\rm c}\sqrt{\gamma_{\rm e}} = 1.382 \; \alpha_{\rm BCS} = 2.438$ [see Fig.~\ref{fig:Relations_alpha-model}~(b)]. The entropy approach gives slightly lower values than the experimental data (black dashed line), whereas the free energy approach, which can be expressed analytically as $B_{\rm c}(0)/T_{\rm c}\sqrt{\gamma_{\rm e}} = 1.382\ \alpha$, overestimates the data (blue dashed line). The different result between the two approaches has already been pointed out in the original $\alpha-$model paper of
Ref.~\onlinecite{Padamsee_JLTP_1973}. The reason is that the strict assumption of a BCS-like temperature evolution of the gap makes the expressions of the free energy and entropy not equivalent for arbitrary values of $\alpha$.

For a better understanding of the experimental results it was further considered the strong coupling corrections of the BCS expressions, which were calculated starting from the Eliashberg equations by various authors.\cite{Carbotte_RMP_1990, Rainer_book_1986, Marsiglio_book_2008, Hake_PR_1967, Rainer_JLTP_1974, Geilikman_JLTP1975, Orlando_PRB_1979, Webb_PhysC_2015} Using the analytical expressions:\cite{Geilikman_JLTP1975, Orlando_PRB_1979}
\begin{equation}
\alpha = \alpha_{\rm BCS} \left[ 1+ 5.3 \left( \frac{T_{\rm c}}{\omega_{0}} \right)^2 \ln \frac{T_{\rm c}}{\omega_{0}} \right] \nonumber
\end{equation}
and 
\begin{equation}
\frac{B_{\rm c}(0)}{T_{\rm c}\sqrt{\gamma_{\rm e}}} = 2.438  \left[ 1+ 2.3 \left( \frac{T_{\rm c}}{\omega_{0}} \right)^2\ln \frac{T_{\rm c}}{\omega_{0}} \right], \nonumber
\end{equation}
where $\omega_{0}$ is a typical phonon frequency of the material, one obtains the simple linear relationship:
\begin{equation}
\frac{B_{\rm c}(0)}{T_{\rm c}\sqrt{\gamma_{\rm e}}}= 1.38 + 0.599 \; \alpha.
\label{eq:Bc_relation}
\end{equation}
The red line in Fig.~\ref{fig:Relations_alpha-model}~(b) represents Eq.~\ref{eq:Bc_relation} and it well reproduces the experimental data without {\it any} adjustable parameters.

It is worth noting that full numerical solutions of the Eliashberg equations tend to agree within 10\% of the experimental data when plotted against the coupling parameter ${T_{\rm c}}/{\omega_{ln}}$ (${\omega_{ln}}$ is the well known logarithmic moment of the spectral phonon density).\cite{Carbotte_RMP_1990} This indicates that  $T_{\rm c}/\omega_{\rm ln}$ can be viewed as hidden parameter
of the universal relations presented in Fig.~\ref{fig:Relations_alpha-model} and that these relations  have their direct consequence also within Eliashberg theory.
The dependence of BCS scaled parameters on  $T_{\rm c}/\omega_{\rm ln}$ was already widely discussed in the literature (see {\it e.g.} Refs.~\onlinecite{Carbotte_RMP_1990, Rainer_book_1986, Marsiglio_book_2008, Hake_PR_1967, Rainer_JLTP_1974, Orlando_PRB_1979, Webb_PhysC_2015}).
On the other hand, our result allows to predict the specific heat jump and the scaled thermodynamical critical field (or viceversa) once the strong coupling ratio is known without resorting to measurements or calculations of the phonon spectra.

The scaling relation between $B_{\rm c}(0)/T_{\rm c}\sqrt{\gamma_{\rm e}}$ and $\Delta/k_{\rm B}T_{\rm c}$ [Fig.~\ref{fig:Relations_alpha-model}~(b) and Eq.~\ref{eq:Bc_relation}] allows an explanation of the empirical relation between the thermodynamic critical field $B_{\rm c}(0)$ and transition temperature $T_{\rm c}$ for type-I superconductors shown in Fig.~\ref{fig:Bc-vs-Tc}.
Indeed, Eq.~\ref{eq:Bc_relation} implies that the ratio $B_{\rm c}(0)/T_{\rm c}$ can be written as:
\begin{equation}
\frac{B_{\rm c}(0)} {T_{\rm c}}= (1.38 + 0.599\; \alpha) \sqrt{\gamma_{\rm e}}.
\label{eq:BcTc-scaling}
\end{equation}
The quantities on the right hand side of Eq.~\ref{eq:BcTc-scaling} do not vary very much among single metal type-I superconductors. For instance
the coupling strength $\alpha$ changes from $\simeq 1.75$ (for Al and Ga-I) up to $\simeq 2.25$ (for Hg and Pb).\cite{Carbotte_RMP_1990, Poole_Book_2014} The spread in the electronic specific heat coefficient $\gamma_{\rm e}$ is also not big. The highest and the lowest $\gamma_{\rm e}$ values are reported for Pb ($\gamma_{\rm e}\simeq 3.10$~mJ/mol K$^2$) and Ga-I ($\gamma_{\rm e}\simeq 0.60$~mJ/mol K$^2$), respectively.\cite{Poole_Book_2014} This implies that $B_{\rm c}(0)/T_{\rm c}$ ratio for various type-I superconductors does not differ by more than a factor of 2-3. This value is much smaller than the range of $T_{\rm c}$ and $B_{\rm c}(0)$ in Fig.~\ref{fig:Bc-vs-Tc}, which spans over almost three orders of magnitude.

To conclude, correlations between coupling strength $\Delta/k_{\rm B} T_{\rm c}$, the ratio $B_{\rm c}(0)/T_{\rm c}\sqrt{\gamma_{\rm e}}$, and the specific heat jump $\Delta C(T_{\rm c})/\gamma_{\rm e} T_{\rm c}$ were found to hold for phonon-mediated superconductors.
The corresponding quantities for the pressure stabilized Ga-II phase, obtained from the temperature dependence of the thermodynamic critical field $B_{\rm c}(T)$ and of the specific heat $C(T)$ follow quite precisely the above mentioned scaling laws. These relations can be well understood taking into account strong coupling adjustments of BCS universal parameters and can naturally explain the empirical scaling
of thermodynamic critical field and critical temperature of phonon-mediated type-I superconductors.

The work was performed at the Swiss Muon Source (S$\mu$S, PSI Villigen, Switzerland).

\end{document}